\documentclass{article}
\usepackage{ijcai18}

\usepackage{times}
\usepackage{soul}
\usepackage{url}
\usepackage{graphicx}
\usepackage{amsmath}
\usepackage{float}
\usepackage[hidelinks]{hyperref}
\usepackage[utf8]{inputenc}
\usepackage[small]{caption}

\title{The Impact of Humanoid Affect Expression on \\ Human Behavior in a Game-Theoretic Setting\thanks{This work was presented at 1st Workshop on Humanizing AI (HAI) at IJCAI'18 in Stockholm, Sweden.}}

\author{
Aaron M. Roth\thanks{These authors contributed equally},
Umang Bhatt\footnotemark[2],
Tamara Amin\footnotemark[2],
Afsaneh Doryab,
Fei Fang,
Manuela Veloso
\\ 
Carnegie Mellon University, Pittsburgh, USA\\
\{aaronr1, usb, tta, adoryab, feif\}@andrew.cmu.edu, mmv@cs.cmu.edu
}

\begin{document}

\maketitle

\begin{abstract}
With the rapid development of robot and other intelligent and autonomous agents, how a human could be influenced by a robot's expressed mood when making decisions becomes a crucial question in human-robot interaction.
In this pilot study, we investigate (1) in what way a robot can express a certain mood to influence a human's decision making behavioral model; (2) how and to what extent the human will be influenced in a game theoretic setting.
More specifically, we create an NLP model to generate sentences that adhere to a specific affective expression profile. We use these sentences for a humanoid robot as it plays a Stackelberg security game against a human. We investigate the behavioral model of the human player. 
\end{abstract}

\section{Introduction}
In the future, robots will be ever-present in our daily lives.  We will live with them, work with them, and engage in all kinds of collaborative and potentially competitive interactions with them. 
Examples of robot sharing workspace with human can already be found today in the realms of elderly care, rehabilitation and healthcare, education, personal companions, and social robots. In order to ensure that robot interact with human in an intended way, it is crucial to gain a better understanding of how a robot affects humans behavior, especially in their decision making process.

Humans evolved to read cues in emotions and moods. The expressions of those in one's surroundings can affect one's own levels of rationality, risk-taking, and decision-making. To what degree does this hold true for a robot companion as well?  After all, a humanoid (or even non-humanoid) machine can act in a manner that humans can perceive as having an ``emotion'' or personality. Social robots, in particular, can take advantage of uniquely human modes of interaction.


In this work, we seek to understand how a robot's affect expression can influence a human's decision making. Thus motivated, we consider a competitive setting where a human will play a Stackelberg security game against a robot. 

We aim to answer two research questions. First, in what way can a robot express a certain mood that can influence a human's decision making behavioral model? To answer this question, we create an NLP model to generate sentences that adhere to a specific affective expression profile in the setting of playing a competitive game. We equip a humanoid robot with verbal feedback system including these sentences. 

Second, how does the observed affect expressed by a humanoid opponent impact a human being’s rationality and strategy in a game theoretic setting? In behavioral game theory, an influential model of human's behavior is the quantal response model~\cite{qre}, and the parameter in the model can be interpreted as the level of rationality of human. We run human subject experiments in a pilot study where we ask the human participants to play a Stackelberg security game~\cite{yang2011improving} multiple times against the affect-expressive humanoid robot. We use Maximum Likelihood Estimation to find the best value of the parameter that fits the data collected from human actions in the game plays, and thus evaluate how the human behavior is impacted. 


\section{Related Work}\label{sec:rw}
\subsection{Affective Expression in Humans and Robots}
Emotions are an important aspect of human interaction. Human beings interpret emotion through nonverbal and verbal cues. In \cite{frijda2005emotion}, Fridja argues that experiencing emotions are, in fact, the primary purpose of social interactions. Notably, it is well documented that the observation of another person's mood can have specific effects on behavior and can influence the observer's mood following the interaction \cite{wild2001emotions}.

Affect is a general term relating to emotions, moods, feelings and other such states. Affective states vary in their degree of \textit{activation} and \textit{valence} (whether they are positive and negative)\cite{22}.  In the psychological and cognitive science literature this is often represented via axes in a continuous multi-dimensional space \cite{44}. Emotion is classified by short term, intense affective states and mood is classified by long term, diffuse states. Emotions are evaluative responses in specific events or to stimuli of importance. Modeling emotion in a robot generally requires the interruption of an interaction, whereas modeling a mood allows for the interaction to persist with only slight differences in the execution of the interaction. For these reasons most studies, including our own, focus on mood and not emotion.

People exchange verbal messages which contain information conveying their mental and emotional states. This includes the use of emotionally colored words and swear words. Given the importance of affect in language, there has been a fairly substantial amount of research in affective statistical language modeling\cite{wagner2014gesture}.
This includes the development of affective NLG for generating medical texts \cite{mahamood2011generating}, and rule based emotive text generation based on sentence patterns \cite{keshtkar2011pattern}. Notably, there has emerged work in the extension of the LSTM (Long Short Term Model) language model for generating affective text \cite{ghosh2017affect}.

Research has shown that human beings are capable of perceiving robotic \textit{affect} under some circumstances. Different forms of affective expression have been modeled with humanoid robots. Bodily expression has been used for ROMAN \cite{55}, NAO \cite{55,56,57}, KOBIAN \cite{54,70}, and Max \cite{99}. These studies demonstrated that people are generally capable of recognizing affective states. Work has also been done to develop a parameterized behavior model in which behavior parameters controlled the spatial and temporal extent of a behavior for mood expression.  This includes models that enable the continuous display of mood in an interactive game \cite{xu2014robot}. Work of this nature with robots is less developed than similar work with software agents, but it is becoming more common.
 However, research focusing on the effect of affective language models in conjunction with human - social robot interaction has been limited (discussed below).

It has been shown that affective expression can have positive effects on humans during an interaction. For example, studies with the robot Vikia have demonstrated the effectiveness of an emotionally expressive graphical face for encouraging interactions with a robot\cite{bruce2002role}. Other effects include the way of interacting with a robot with emotive facial expressions \cite{42}, the effectiveness of assistive tasks such as learning and motivation given vocal emotion expression\cite{46}, user behavior during support tasks \cite{47}, and user mood. Of particularly interest to us are studies in the format of interactive games and competitions. 

Mood contagion is a well researched automatic mechanism whereby the observation of another person's emotional expression induces a congruent mood state  in the observer. Prior research has also identified that body language of a robot is contagious in an 
imitation game setting \cite{xu2014robot}.

Several studies also reported effects of affective virtual agents user task performance. Experiments with GRETA showed that consistent emotion expression resulted in better performance of recall \cite{x106}. Emotion expression was reported to have effects on users' affective states and behaviors. As of yet there are has been little research on the impact of observed robotic affect on user performance  

The majority of studies so far utilizing games have focused on learning games such (e.g. memory and imitation). Currently, mood contagion is an area that is receiving a great deal of interest, and although there still exists research on performance, it focuses on outcomes and is largely based on virtual agents. To date, there have been \textbf{no studies}, that we know of, which explore the influence of robotic affect on a player's strategy, risk-taking, or rationality.

\subsection{Quantal Response}
Quantal Response is a technique for extracting human rationality \cite{qre}. We want to understand the error in an individual's response and understand the probability distribution of the possible responses.  If there are multiple actions in a situation, and the utility of a given action $i$ is $U_i$, then the best quantal response $q_i$ is calculated as follows, where $\lambda$ is used to indicate the deviation from optimality:
\begin{equation}\label{eqn:qr0}
q_i =  \frac{e^{\lambda U_i}}{\sum_{j=1}^{n} e^{\lambda U_j}}
\end{equation}
We extend the work of \cite{yang2011improving} wherein the best response quantal response is used to figure out how rational a player is in a Stackelberg Security game. We hold the defending strategy constant but aim to model the $\lambda$ of each human player of our game.  In this formulation, $U_i$ in equation \ref{eqn:qr0} represents the utility of selecting the $i^{th}$ gate in a given round. Using maximum likelihood estimation, we can find the most likely value of $\lambda$ for a given player given the utilities in the given round and the gate choices of the player over the all rounds in the game. We can then compare these $\lambda$ values over different players to see which players are more rational. A $\lambda$ of 0 would indicate `irrational' behavior (selecting actions in a uniform random manner), whilst a higher $\lambda$ would indicate a higher human rationality (selecting more `optimal' actions).  

\section{Our Approach}
\begin{figure}
\includegraphics[width=\columnwidth]{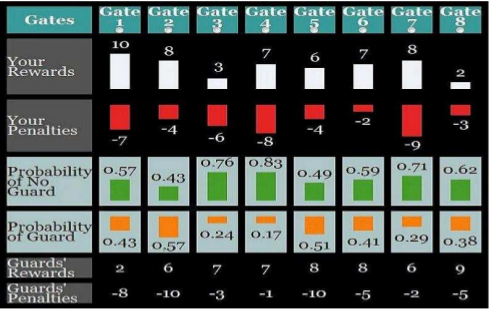}
   \caption{The gate game.}
   \label{gate-pic}
\end{figure}

We run human-subject experiments.
Participants will be presented with and play the game shown in figure \ref{gate-pic}.  We based this on the gate game found in  \cite{yang2011improving}. \footnote{This game is in the public domain and at the time of writing, can be downloaded from \url{http://teamcore.usc.edu/Software.htm}}  A series of ``gates'' is presented which the human can ``attack.''
First, they play against a computer opponent.  (This serves as a baseline.) Then, they will play ``against'' a humanoid robot (a Softbank Pepper robot).
When they play against the humanoid, the robot will act in either an encouraging or discouraging manner.  This will be achieved by word and sentence usage generated by an NLP model we have created.
We record the gate information with which they are presented and their choices, and from this we can calculate their quantal response.

\subsection{Methods}

\subsubsection{N-gram Sentence Predictor for Affective Phrases}

The probability of a sequence of words is found using the Chain Rule of Probability
\begin{equation}\label{eqn:chain-rule}
P(w_1^n) = P(w_1) P(w_2| w_1) P(w_3| w_1^2)...P(w_n|w_1^{n-1})
\end{equation}
\begin{equation}\label{eqn:chain-rule-product}
P(w_1^n)= \displaystyle\prod_{k=1}^{n}P(w_k|w_1^{k-1})
\end{equation}

We will utilize the Markov assumption which estimates that the probability of some future event depends only on a limited history of preceding events. Based on this assumption we developed an N-gram Model which approximates the probability of a word by looking N-1 words into the past.
\begin{equation}\label{eqn:n-gram}
P(w_n|w_1^{n-1}) \approx P(w_n| w_{n-N+1}^{n-1})
\end{equation}
The probability of a word is estimated using Maximum Likelihood Estimate by normalizing n-gram counts to obtain relative frequency.
\begin{equation}\label{eqn:ngram-rel-freq}
P(w_n| w_{n-N+1}^{n-1}) = \frac{C(w_{n-N+1}^{n-1} w_n)}{C(w_{n-N+1}^{n-1})}
\end{equation}
To account for unseen events we utilize Laplacian (add-one) smoothening
\begin{equation}\label{eqn:ngram-rel-freq}
P(w_n| w_{n-N+1}^{n-1}) = \frac{C(w_{n-N+1}^{n-1} w_n)+ \alpha}{ C(w_{n-N+1}^{n-1})+ \alpha{D}}
\end{equation}
where $\alpha$=1 and $D$ is the length of the vocabulary dictionary.

\textbf{Bidirectional Predictor:} To develop a fill-in-the-blank model we created Trigram (N=3) and Bigram (N=2) models that operated in both the forward and backward direction(total of four models). The forward models are called to predict the blank based on the words preceding it and the reverse models are called to the predict the blank based on the word following it. 

\textbf{Corpora:} For the training process the following corpora were used from the python nltk set of corpora: brown, gutenberg, inaugural, state of the union, and genesis (English text). These texts were tokenized into sentences, made case not-sensitive and eliminated of punctuation. For the reverse model,  the corpora (and the words following the blank) are reversed prior to modeling and prediction. 

\textbf{Hierarchy:} If there were no examples of a particular trigram,$(w_{n-2}, w_{n-1}, w_n)$, to compute $P(w_n|w_{n-2}, w_{n-1})$, we can estimate its probability by using the bigram probability $P(w_n|w_{n-1} )$.

\textbf{Filtering:} Once the probabilities for predicted words are obtained for each model, the results are filtered to eliminate stop words and numerals.

\textbf{Affect Weighting:} To create emotive phrases we utilized the AFINN Affect Dictionary which provides a list of English words that are rated based on emotional `valence' (assigned values from -5 to +5) \cite{nielsen2011new}. For our procedure, each phrase fed to the model yields two sets of predictions - corresponding to positive emotion assignment and negative emotion assignment. The desired affective state is codified by a binary assignment ($0 = negative$, $1 = positive$). Word predictions are filtered using the AFINN data such that only words with the desired emotional valence are included. An Affect Score for each prediction is then calculated based on the strength of the assigned emotional valence. This is calculated as a fraction ranging from ${|\frac{1}{5}|}$ to ${|\frac{5}{5}|}$.

\textbf{Finite Mixture Model:} To combine the probability estimators, we constructed a linear combination of the multiple probability estimates and weighted each contribution so that the result is another probability function. We weighted each contribution equally ($\lambda{=0.2}$ and included a contribution $A$ corresponding to the Affect Factor discussed above).
\begin{equation}
\label{eqn:chain-rule-product}
\begin{aligned}
P(w_n|w_{n-2}w_{n-1}) =  &  \lambda{P_f(w_n|w_{n-2}, w_{n-1})} \\
 &+ \lambda{P_r(w_n|w_{n-2}, w_{n-1})} \\
 &+ \lambda{P_f(w_n|w_{n-1})}\\
 &+ \lambda{P_r(w_n|w_{n-1})} + \lambda{A}
\end{aligned}
\end{equation}
where the subscript $r$ corresponds to the reverse direction and the subscript $f$ corresponds to the forward direction.

\textbf{Final Manual Filtering:} Our algorithm presents the two highest scoring words to the end user. As a final step, we manually eliminated any words that were grammatically incorrect, illogical in the sentence context or inappropriate. Following this, if the remaining highest score was shared by numerous words, we manually selected one based on preference or words previously unseen. Words with the second highest score were selected if none of the highest scoring words survived filtering.


\subsubsection{General Quantal Response}\par
As shown in figure \ref{gate-pic}, the human can choose from among eight gates to attack.  They attack one gate out of the eight.  The reward and penalties for each gate are shown to the user.  If they attack a gate and it is defended, they incur a penalty, and if they attack a gate and it is not defended, then they receive the reward.  The human will have their own probably imperfect strategy, but we can calculate the mathematical expected utility $G_j$ for a given gate $j$ as follows:

\begin{equation}
G_j = (1 - p_j) R_j + p_j P_j
\end{equation} 
where $p_j$ is probability of a guard at gate $j$, $R_j$ is the attacker's reward at gate $j$, and $P_j$ is the attacker's penalty at gate $j$. We can expand the work from \cite{yang2011improving}. We calculate the log likelihood of the quantal response for a given gate $i$ in round ($r$) for a given player as follows. 
\begin{equation}
\log(q_{ri}) = \lambda G_{ri} - \log(\sum_{j=1}^{N}{\exp(\lambda G_{rj})})
\label{mle}
\end{equation}
We sum over all possible gates that the player could have chosen in the round $r$. Our goal is to learn the $\lambda$ of the player to better understand that player's rationality. The higher the $\lambda$, the more rational the player. We posit that rationality (that is, a player's $\lambda$ value) will differ depending on what affect the humanoid exhibits. We want to find the maximum likelihood estimate of $\lambda$; we can do this by using a Mixed Integer Linear Program (MILP) to set up maximizing equation \ref{mle} as an objective with the constraint that $\lambda$ is nonnegative.

\subsubsection{Emotion-Parameterized Quantal Response}
Extending the work of \cite{TambeHumanAI}, we propose a novel approach to Subjective Utility Quantal Response, which we hereafter call Emotion-Parameterized Quantal Response (EPQR). We show that we can create a input vector $X$ which would capture the reward, penalty, and the emotional affect. In our case, $X$ can be formalized as follows: $X_j = [R_j, P_j, E_j]$ where $E_j$ is 0 when the humanoid is in the negative affect class and 1 if in the positive affect class and where $R_j$ and $P_j$ are the reward and penalty of the $j^{th}$ gate in the given round. We can also explicit encode for both emotions by tacking on an additional $4^{th}$ term to $X_j$. This latter technique has the added benefit of learning parameter weights for both emotions independently. If we had not made the assumption that the two chosen affect classes are complements (positive and negative verbiage), we would prefer to learn an explicit weight for every emotion, as a human can vary in their sensitivity to different emotions. We can formalize the log likelihood of EPQR for a given round as follows, where the vector $w$ captures the weights of the elements in $X$ over all rounds.

\begin{equation}
\log(eq_{ri}) = w^{T} x_{ri} - \log(\sum_{j=1}^{N}{\exp(w^{T} x_{rj}})
\end{equation}
\begin{equation}
w^{T} x = w_1 x_1 + w_2 x_2 + w_3 x_3
\end{equation}

\subsection{Experimental Design}

This section describes the procedure for running an experiment for a test subject.

\subsubsection{Consent}
This consent form explains the purpose of the study, and gives a summary of what the participant can expect to experience.
There are no risks aside from breach of confidentiality. They are informed that video and audio recordings may be made. We do not tell participants that we are measuring their mood, analyzing their game strategy, or learning about their reaction to the robot.

\subsubsection{Pre-Game Survey}
We administered a written survey before they saw the gate game or the robot.  It measured their starting emotional state. 

\subsubsection{Practice Rounds}
The participant was given the iPad and played practice rounds of the game ``against the computer.''  In this manner they learned how to play before encountering the robot.  We can also measure their ``baseline'' quantal response unaffected by an opponent's emotions as the control data.

\subsubsection{Repeated Games Against the Robot}
The participant would then be led into the room where the robot was located.  They would sit across from the robot with an Ipad between them.  A video camera would be fixed upon them. The researchers would be hidden from view by a screen, so that the participant and the robot had some privacy, and the presence of the researchers would not affect the participant.\footnote{During the feasibility study, no video camera was used.  During the feasibility study, the researchers were not hidden by a screen.} The participant would play a series of rounds of the game ``against'' the robot.  During play, the robot would comment on the game and on the participant. The robot would exhibit one of two opposing affects--either the positive (encouraging) affect or the negative (discouraging) affect.  The only practical difference was in the particular word choice used by the robot when it spoke to the participant. Although the robot would describe the game as if it had some understanding of what was going on, in reality the participants' actual performance had no effect on what the robot said.  Encouragement and discouragement and comments on how well or poorly the person was playing were the same or similar within a particular affect category, regardless of whether the participant was generally winning or losing.

\subsubsection{Post-Game Surveys and Video}
After the game concludes, a researcher re-enters and gives the participant a written survey to fill out. Then, the participant is asked a series of questions verbally by the researcher, to which they respond on the video camera.\footnote{We encrypt the data with a NIST-approved encryption scheme. Participants are assigned ID numbers and data is labeled according to ID numbers rather than names.}

\subsubsection{Debriefing}
In the initial stages, participants were told they would play a game against a robot but were not told the true purpose of the study or that the robot was expected to have a particular affect. Now, we explained to them that we were studying how their strategy changed depending on the emotions of the robot, and their perception of the robot thereby. We would explain why this deception (of not telling the participant about the humanoid's affect expression) was necessary.

\section{Empirical Results}
\subsection{Natural Language Sentence Generation}
The NLP sentence generation via fill-in-the-blanks with emotion-weighted trained n-gram predictions yielded excellent results.  Some sample sentences generated are shown in figure 2.

\begin{figure*}[t]
\begin{tabular}[c]{l|l}
\textbf{Positive Sentences} & \textbf{Negative Sentences} \\
\hline
Honestly, this game is a successful experience. & Honestly, this game is a disastrous experience. \\
I can tell you have a very great personality. & I can tell you have a very selfish personality. \\
I hope that you are having a wonderful time. & I hope that you are having a challenging time. \\
I am starting to get a sense of your expert strategy. & I am starting to get a sense of your ridiculous strategy \\
Over the course of this game your playing has become great. & Over the course of this game your playing has become confused.
\end{tabular}
\caption{Example Sentences Built by Our NLP Algorithm}
\label{tbl:ngram_sentences}
\end{figure*}

The feasibility study seemed to validate that the sentences built in this manner did indeed convey the desired affect to the participant.
Each participant in the feasibility study was asked to describe their perceptions of the robot in a written post-game survey.
The participant who experienced the discouraging personality described the robot as ``very discouraging'' and even wrote that the robot ``made me question myself.''
In contrast, the participant who experienced the encouraging personality described the robot as ``very encouraging'' and wrote, ``I like him.''
It is gratifying to us that the very words we used to describe the opposing affects in our development of the study showed up in the participants' answers to a free-response question. It suggests the validity of our NLP model.

\subsection{Rationality Analysis}
We extract the rationality of the two subjects in our pilot study.  We find the $\lambda$ value that captures their ability to play the optimal strategy against the humanoid. We can find the maximum likelihood estimate of $\lambda$ of each player. Below we plotted the cumulative $\lambda$ for two players (each shown a different affect class) over the first 8 rounds. (We calculate $\lambda$ after every round using all of the data from the past rounds and current round.)

\begin{figure}[H]
\includegraphics[width=\columnwidth]{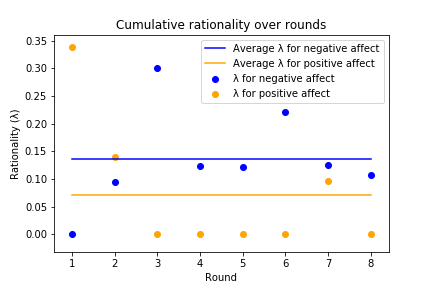}
   \caption{Cumulative rationality of two players over 8 rounds of play}
   \label{foo}
\end{figure}

Contrary to what we expected, we found that the player exposed to the negative verbiage was more rational than the player who was encouraged by the humanoid. Over all thirty five rounds, we found that the player exposed to the negative affect class had a final cumulative $\lambda$ of $0.128$ while the player shown the positive affect class had a a final cumulative $\lambda$ of $6 \times 10^{-9}$. The positive player's behavior was almost uniform random, which is the highest level of irrationality on our scale. The negative player became quite rational over time.  Although no conclusions can yet be drawn, it could be the case that the humanoid's expression of negative affect drives up human rationality.  We hope to investigate this initial result with further rigorous experimentation.   It is also possible that the data we have collected so far are anomalies, in which case future work would reveal different trends. 

\section{Future Work}




We observed size of the affect-assigned dataset, and not the size of the training corpora, that was the primary limitation to obtaining diverse final predictions. Future work could utilize the more robust LIWC (Language Inquiry and Word Count), however, this would prevent the algorithm from being fully open source. 

While the N-gram model is effective as a simple means of generating language, other more robust methods should be explored, such as sentence pattern recognition and the use of deep learning as outlined by Ghosh et. at.\cite{ghosh2017affect}.

\section{Conclusions}

We have developed a language model that seems to cause a humanoid robot to be perceived with the intended affective mood, for the affects of ``Encouraging'' and ``Discouraging''.  The sentences used in the two conversations are similar, since they are simply fill-in-the-blank sentence stems.  The entirety of the rest of robot -- voice intonation, positioning, appearance -- was constant for all participants.  This shows how choosing the appropriate word, when the rest of the sentence is neutral, can have a significant impact on the perception of the agent overall.  The feasibility study suggests that our language model is simple yet powerful, working as well as we had desired, and suitable for our intended purposes.

After analyzing the quantal response of our two test subjects, our preliminary results show that negative affect leads to a higher $\lambda$ for a player, meaning they are comparatively more rational than a player who was shown the positive affect.  This would seem to be a surprising result if it is not an anomaly (essentially, discouragement yielded better playing by the discourager's opponent), and so indicates that this avenue of research is  worth pursuing.

This study is ongoing. Eventually we will have a wealth of data on the impact on a human's decision-making when engaged in these types of human-robot interactions in a game-theoretic situation.

We hope this research, as it continues, will inform future researchers and designers in developing social robots whose attitudes and emotional reactions and responses are conducive to productive social human-robot interactions.

\section*{Acknowledgments}
We would like to thank the following researchers for their guidance and/or advice: Jeffery Cohn (Robotics Institute/Department of Psychology, CMU/Pitt), Louis Philippe Morency (Language Technologies Institute, CMU), and Samantha Reig (Human-Computer Interaction, CMU).  

\bibliographystyle{named}
\bibliography{ijcai18}

\end{document}